\documentclass[a4paper,11pt]{article}
\pdfoutput=1

\usepackage{slashed}
\usepackage{amsmath}
\usepackage{amssymb}
\usepackage{amsthm}
\usepackage{indentfirst}
\usepackage{psfrag}
\usepackage{graphicx}
\usepackage[utf8]{inputenc}

\usepackage{jheppub}
\begin{document}

\title{Coset space construction for the conformal group. II. Spontaneously broken phase and inverse Higgs phenomenon.}

\author[a,b]{I. Kharuk}

\affiliation[a]{Moscow Institute of Physics and Technology,\\
Institutsky lane 9, Dolgoprudny, Moscow region, 141700, Russia}
\affiliation[b]{Institute for Nuclear Research of the Russian Academy of Sciences,
\\ 60th October Anniversary Prospect, 7a, Moscow, 117312, Russia}

\emailAdd{ivan.kharuk@phystech.edu}

\preprint{INR-TH-2017-032}

\abstract{
A mathematically strict method of obtaining effective theories resulting from the spontaneous breaking of conformal invariance is developed. It is demonstrated that the Nambu--Goldstone fields for special conformal transformations always represent non--dynamical degrees of freedom. For spacetime manifolds with dimension $ d>2 $, the equivalence of the developed approach and the standard one, which includes the imposition of the inverse Higgs constraints, is proved. Based on the consideration of the special case $ d=2 $, the extension of the technique used in the inverse Higgs phenomenon is discussed.}

\maketitle

\section{Introduction}
\label{sec:1}

The study of effective theories resulting from the spontaneous symmetry breaking of the conformal invariance\footnote{By definition, the conformal group in $ (d+1) $ spacetime dimensions is $ O(2,d+1) $. Not to be confused with the Weyl group, which consists of arbitrary local rescalings of the metric.} has a long history. In the pioneering work \cite{Salam:1970qk} it was noticed that the straightforward application of the coset space technique in this case yields the Nambu--Goldstone fields (NGF) for special conformal transformations (SCT) to be massive. This was suggested to be the manifestation of Anderson--Higgs--Kibble mechanism \cite{anderson1963plasmons,higgs1964broken,guralnik1964global}, which gives rise to a massive vector. However, in the later studies \cite{Isham:1970gz,Volkov:1973vd,Low:2001bw} it was shown that the NGF for SCT do not correspond to the independent fluctuations of the vacuum, whatever the latter is, hence represent redundant fields. An analogous phenomenon was also observed for other spontaneous symmetry breaking patterns \cite{Ivanov:1975zq,Nicolis:2013sga,Watanabe:2013iia}. The way of obtaining a theory with the correct number of degrees of freedom, i.e., without redundant fields, was proposed in \cite{Ivanov:1975zq}. Namely, it was suggested to impose the so called inverse Higgs constraints, which allow to express redundant fields in terms of the physical ones. Such prescription turned out to be successful in all known examples, and, therefore, was accepted as the correct tool for eliminating unphysical fields from a theory. For the spontaneous breaking of the conformal group, this construction allows to eliminate the NGF for SCT in favour of the dilaton field \cite{Ivanov:1975zq,Hinterbichler:2012mv,Volkov:1973vd} and is accepted as the standard approach to the construction of effective Lagrangians.

Note, however, that the mathematical status of the standard approach remains of a proposal. Namely, imposing inverse Higgs constraints is a mathematically consistent way of realizing a given spontaneous symmetry breaking pattern by a smaller number of NGF than there are broken generators. Nonetheless, the necessity of imposing inverse Higgs constraints does not follow from the logic of the coset space technique by itself, which turns the standard approach into a successful \textit{ad hoc} prescription. Thus, the question of how to justify mathematically the standard construction remains open.

Until recently, there was a similar problem with elaborating the way how the coset  space technique should be employed for the construction of conformally invariant Lagrangians in the unbroken phase.\footnote{We say that a theory is in the unbroken phase if none of the symmetries are spontaneously broken, i.e., all symmetry generators annihilate the vacuum. In \cite{Kharuk:2017jwe}, all of the generators were assumed to be unbroken, and the construction of conformally invariant theories was made along the lines of the method of induced representations \cite{Mackey:1969vt,hermann1966lie,Ortin:2015hya}. The presence of the exponentials of the unbroken SCT in the coset space employed in \cite{Kharuk:2017jwe} was necessitated by geometrical considerations.} Namely, to make the coset space technique applicable in this case,\footnote{Coset space technique is applicable only when coset space $ G/H $ is homogeneously reductive. To ensure this property, one should add the exponentials of the SCT into the coset. We remind a reader that a coset space $ G/H $ is said to be homogeneously reductive if $ [Z,V] \subset Z $ and $ [V,V] \subset V $, where $ V $ are generators of $ H $ and $ Z $ supplement them to the full set of $ G $'s generators.} one must consider SCT as if they were broken (more precisely, as non--linearly realized generators) \cite{Salam:1970qk,Wehner:2001dr,Ivanov:1981wm}. This raises the question of how to interpret the coordinates associated with SCT in the corresponding coset space. More concretely, the suitable pattern of non--linear realization reads\footnote{When employing coset space technique, one should specify non--linearly realized generators. Since such generators are not necessarily broken, it is more appropriate to speak of patterns of non--linear realization rather than spontaneous symmetry breaking patterns. For example, even if translations are not spontaneously broken, one must still add them to the coset space \cite{Ogievetsky1974}. Another example covers the construction of gauge theories \cite{Ivanov:1976zq,Goon:2014ika}.}
\begin{equation}
\text{Conf}(d) \rightarrow SO(d) \;,
\end{equation}
where $ \text{Conf}(d) $ is the $ d $--dimensional conformal group.\footnote{For the simplicity reasons (see footnote \ref{footOnEucl} below), we consider the Euclidean conformal group.} This corresponds to considering the coset space
\begin{equation} \label{Conf_Coset}
g_{conf} = e^{iP_\mu x^\mu} e^{iK_\nu y^\nu} \;,
\end{equation}  
where $ P_\mu $ and $ K_\nu $ are generators of translations and SCT accordingly. 
The interpretation of $ x^\mu $ in this coset is known --- they are coordinates on the Minkowski spacetime. But how should one interpret $ y^\nu $? Should they be considered as NGF \cite{Salam:1970qk} or as a set of additional coordinates, thus leading to the so called bi--conformal space \cite{Wehner:2001dr,Ivanov:1981wm}? In \cite{Kharuk:2017jwe}, based on the method of induced representations \cite{Mackey:1969vt,hermann1966lie,Ortin:2015hya}, it was shown that $ y^\nu $ plays a special role. Namely, $ y^\nu $ should be considered as field whose dependence on the coordinates is fixed by the symmetries. This implies that $ y^\nu $, the NGF for SCT, can play a special role in the case of spontaneously broken conformal group as well. If so, this observation can be of crucial importance for justifying the standard approach to the construction of effective Lagrangians resulting from the spontaneous breaking of the conformal symmetry.   

The aim of this paper is to show that this is indeed the case. For this purpose, we extend the the area of applicability of the technique developed in \cite{Kharuk:2017jwe} to a spontaneously broken phase, i.e, allow some of the generators to be spontaneously broken. This allows us to show that in the spontaneously broken phase $ y^\nu $ has fixed dependence on the coordinates, which forces $ y^\nu $ to decouple from other fields. The procedure of fulfilling this requirement is elaborated and the connection with the standard approach is discussed. It is shown that the standard approach can be considered only as a convenient tool for constructing effective Lagrangians, but not as a mathematically self--contained method. In the special case when spacetime dimension $ d $ equals two, it is demonstrated that the application of the standard technique is problematic, since it involves irremovable ambiguities. On the other hand, the new technique allows one to systematically construct effective Lagrangians. Based on this observation, an extension of the standard technique is suggested.

The paper is organised as follows. In section \ref{sec:2}, the standard approach is reviewed and the technique developed in \cite{Kharuk:2017jwe} is generalized to a spontaneously broken phase. In section \ref{sec:3}, it is shown that in $ d>2 $ any effective Lagrangian obtained in the developed approach can be obtained using the standard technique as well. This allows us to reinterpret the standard technique as a convenient tool for the construction of effective Lagrangians. In section \ref{sec:3-2}, we discuss the special case $ d=2 $ and the question of how the standard technique should be extended to become applicable in this case. Finally, section \ref{sec:4} concludes the paper.

\section{Spontaneously broken conformal group}
\label{sec:2}

\subsection{Standard technique}
\label{sec:2-1}

We start by reviewing the standard approach to the construction of effective Lagrangians resulting from the spontaneous breaking of the conformal symmetry. We do not overview the standard rules of applying the coset space technique to the construction of effective Lagrangians, which can be found in \cite{Hinterbichler:2012mv,Ogievetsky1974,Weinberg:1996kr}.  

Suppose that the conformal group was spontaneously broken down to the Poincare subgroup by non--zero expectation value of an order parameter $ \Phi $. This accounts for considering the following pattern of non--linear realization,
\begin{equation} \label{IndRepPattern}
\text{Conf}(d) ~ \rightarrow ~ SO(d) \;.
\end{equation}
The coset space associated with this pattern reads
\begin{equation} \label{CosetBroken}
g_{br} = e^{iP_\mu x^\mu} e^{iK_\nu y^\nu(x)} e^{iD \pi(x)} \;,
\end{equation}
where $ D $ is the generator of dilations. Following the standard rules of applying coset space technique in the case of spontaneous breaking of spacetime symmetries \cite{Ogievetsky1974}, one interprets $ y^\nu(x) $ and $ \pi(x) $ as the NGF, and $ x^\mu $ as the coordinates of the Euclidean space.

To see that $ y^\nu $ is, in fact, a redundant field, consider the action of a dilation and SCT on the order parameter,
\begin{equation}
\hat{D} \Phi = \Delta_\Phi \Phi \,, ~~~ \hat{K}_\mu \Phi = 2x_\mu \Delta_\Phi \Phi \;,
\end{equation}
where $ \Delta_\Phi $ is $ \Phi $'s scaling dimension and it was taken into account that $ \Phi $ does not depend on the coordinates. This formula shows that the action of SCT reduces to the coordinate dependent action of dilations. Or, in other words, SCT do not have their ``own'' action on fields. Then, since NGF are local fluctuations of the vacuum, one does not need $ y^\nu $ to describe all possible fluctuations of $ \Phi $ \cite{Low:2001bw,Ivanov:1975zq,Nicolis:2013sga,Watanabe:2013iia}. Hence, physical consideration shows that $ y^\nu $ represents a redundant field.

To account for this fact in the coset space technique approach, one imposes the so called inverse Higgs constraints \cite{Ivanov:1975zq}. This goes as follows. The Maurer--Cartan forms for coset space (\ref{CosetBroken}),
\begin{equation}
g_H^{-1} d g_H = i P_\mu \omega_P^\mu + i K_\nu \omega_K^\nu + i D \omega_D + i L_{\mu\nu} \omega_L^{\mu\nu} \;,
\end{equation}
where $ L_{\mu\nu} $ are the generators of the Lorentz transformations, are given by
\begin{equation} \label{MCF}
\begin{gathered} 
\omega^\nu_K = e^{-\pi} ( dy^\nu + 2y_\rho dx^\rho y^\nu - y^2 dx^\nu )\,, ~~~~ \omega^\mu_P = e^{\pi} dx^\mu\,, \\ \omega_D = 2y_\rho dx^\rho + d\pi\,, ~~~~ \omega_L^{\mu\nu} = y^\nu dx^\mu - y^\mu dx^\nu \;.
\end{gathered}
\end{equation}
All of them, except for $ \omega_L^{\mu\nu} $, transform homogeneously under the action of the continuous elements of the conformal group. Because of this fact, it seems to be consistent with the symmetries to set $ \omega_D $ to zero,
\begin{equation} \label{StandardIHC}
\omega_D = 0 ~~~ \Rightarrow ~~~ y_\nu = - \frac{1}{2} \partial_\nu \pi \;.
\end{equation}
This prescription is known as the inverse Higgs constraint and, as indicated, allows one to express $ y^\nu $ in terms of the dilaton in a way compatible with all continuous symmetries. As we will see later, however, this prescription is not compatible with the discrete inversion symmetry.\footnote{To define the inversion $ I $ strictly, consider $ O(1,d+1) $ as a symmetry group of hypersurface $ -p^2_0 + p_1^2 + ... + p_{d+1}^2 = 0 \subset \mathbb{R}^{1,d+1} $. Then, $ I $ is the element of $ O(1,d+1) $ changing the sign of $ p_0 $.} Let us, nonetheless, proceed further. By using the left Maurer--Cartan forms with $ y^\nu $ given by (\ref{StandardIHC}), one can construct effective Lagrangians with the correct number of degrees of freedom. Namely, by using the standard rules \cite{Ogievetsky1974}, from Maurer--Cartan forms (\ref{MCF}) one reads out the tetrads $ e^\mu_\nu $, the covariant metric $ g_{\mu\nu} $, and the covariant derivative of $ y^\nu $ (which, upon imposing inverse Higgs constraint (\ref{StandardIHC}), plays the role of dilaton's covariant derivative),
\begin{equation} \label{CovarDerDilatIHC}
\begin{gathered} 
\omega_P^\mu = e^\mu_\nu dx^\nu \,, ~~~ g_{\mu\nu} \equiv e^\lambda_\mu \delta_{\lambda\rho} e^\rho_\nu = e^{2\pi}\eta_{\mu\nu} \,, \\ 
D_\mu y^\nu|_{ihc} = e^{-2\pi} \Big( \frac{1}{2}\partial_\mu \pi \partial^\nu \pi - \frac{1}{2} \partial_\mu \partial^\nu \pi - \delta^\nu_\mu \partial_\lambda\pi \partial^\lambda \pi \Big) \;.
\end{gathered}
\end{equation} 
For a matter field $\psi $, one can introduce the homogeneously transforming 1--form \cite{Ogievetsky1974,Weinberg:1996kr}
\begin{equation}
D\psi = d\psi + \omega^{\mu\nu}_L \hat{L}_{\mu\nu} \psi \;,
\end{equation}
where $ \hat{L}_{\mu\nu} $ is a representation of $ L_{\mu\nu} $ appropriate for $ \psi $. On inverse Higgs constraint (\ref{StandardIHC}), this translates to the following covariant derivative,
\begin{equation} \label{CovarDerMatterIHC}
D_\mu \psi |_{ihc} = e^{-\pi} \Big( \partial_\mu \psi + \frac{1}{2} \big( \delta^\rho_\mu \partial^\lambda \pi - \delta^\lambda_\mu \partial^\rho \pi \big) \hat{L}_{\lambda\rho} \psi \Big) \;.
\end{equation}
Then, any effective Lagrangians constructed from (\ref{CovarDerDilatIHC}) and (\ref{CovarDerMatterIHC}) in a $ SO(d) $--invariant way will automatically be conformally invariant as well. Moreover, the effective Lagrangians will include only one NGF, the dilaton, as it was required by the physical considerations.

The presented technique constitutes the standard approach to the construction of effective Lagrangians resulting from the spontaneous breaking of the conformal symmetry. Despite its success, the requirement (\ref{StandardIHC}) remains its weak point. On the one hand, one must impose it to obtain theories with the correct number of degrees of freedom. On the other hand, from the coset space technique it is unclear why NGF for SCT are always redundant. For example, based only on mathematical aspects, one can assume that in some cases $ y^\nu(x) $ will indeed be a massive vector NGF or will have some other physical interpretation \cite{Isham:1970gz}.

\subsection{Two--orbit approach}
\label{sec:2-2}

Before extending the area of applicability of the technique developed in \cite{Kharuk:2017jwe} to a spontaneously broken conformal group, let us remind the reader how it works in the unbroken phase. Throughout the paper, this technique will be referred to as the two--orbit approach.

The fields of $d$-dimensional conformal field theories are defined on a sphere $ S^d $, which is equivalent to the Euclidean space supplemented by a point at infinity.\footnote{\label{footOnEucl}In the Minkowski spacetime, one would have a ``light cone'' at infinity. This is the reason why we are considering Euclidean conformal group.} This is so because SCT always map some point to the infinity, thus forcing one to include it into consideration. Hence, according to the use of the coset space technique in the method of induced representations \cite{Mackey:1969vt,hermann1966lie,Ortin:2015hya}, for the construction of conformally invariant Lagrangians in the unbroken phase one should find the coset space isomorphic to $ S^d $. At this point, one encounters a difficulty. Namely, on the one hand, since in the process of obtaining Maurer--Cartan forms one takes a logarithmic derivative of a coset, the coset to be used must be parameterized by continuous group elements. On the other hand, a sphere is \textit{not} isomorphic to an orbit of any of its points under the action of \textit{continuous} group elements (remember that only the inversion maps the south pole to the north pole). To resolve this difficulty, one may exploit the fact that a sphere can be obtained by gluing together two Euclidean spaces. A rigorous implementation of this idea leads to the following technique \cite{Kharuk:2017jwe}: for the construction of conformally invariant theories, one should consider the coset space corresponding to pattern of non--linear realization (\ref{IndRepPattern}), (\ref{Conf_Coset}), in which $ y^\nu $ is considered as a field with the fixed coordinate dependence,
\begin{equation} \label{Glue}
y^\nu (x) = \frac{x^\nu}{x^2}\,, ~~\vec{x} \neq \vec{0}\;.
\end{equation}
$ y^\nu $ describes the gluing map of coordinate charts around the south and north poles of the sphere, and thus turns a $ 2d $ dimensional coset space (\ref{Conf_Coset}) into a $ d $ dimensional sphere. Note also that condition (\ref{Glue}) is a consistent requirement in the sense that it is invariant under the action of the conformal group. Indeed, eq. (\ref{Glue}) is a solution of the equation $ \omega_K^\nu = 0 $ \cite{Kharuk:2017jwe}. Then, since $ \omega_K^\nu $ transforms homogeneously under the action of the conformal group, eq. (\ref{Glue}) is invariant.\footnote{The action of the inversion interchanges the roles of $ x^\mu $ and $ y^\nu $, and eq. (\ref{Glue}) respects this symmetry as well \cite{Kharuk:2017jwe}. On the other hand, the second solution of the equation $ \omega_K^\nu = 0 $, namely $ y^\nu = 0 $, is not compatible with the action of the inversion.  Also, it is worth stressing that, unlike the approach adopted in \cite{Ivanov:1981wm}, we consider $ y^\nu(x) $ as a field, not as a coordinate.} 

Condition (\ref{Glue}) univocally fixes $ y^\nu $ as a function of the coordinates, hence $ y^\nu(x) $'s equation of motion must admit (\ref{Glue}) as a solution. This fact strongly constrains the allowed combinations of the Maurer--Cartan forms for coset space (\ref{Conf_Coset}). Namely, according to the coset space technique, a covariant derivative of a (quasi--primary) field $ \psi $ reads
\begin{equation} \label{CovarDerField}
D_\mu \psi = \partial_\mu \psi + 2y^\nu \left( \eta_{\nu\mu}\Delta +i \hat{L}_{\mu\nu} \right) \psi \;,
\end{equation}
where $ \Delta $ and $ \hat{L}_{\mu\nu} $ are, accordingly, representations of $ D $ and $ L_{\mu\nu} $ appropriate for $ \psi $. Then, on the one hand, $ y^\nu $ must be given by (\ref{Glue}). On the other hand, $ \psi $, which is a dynamical field, couples to $ y^\nu $ via the interaction term in (\ref{CovarDerField}). This will force $ y^\nu $ to be a dynamical as well, thus spoiling the first requirement. Hence, the only way to avoid this problem is to require for the interaction terms to sum up to a total derivative or disappear at all. This is a qualitatively new requirement one encounters in the process of constructing conformally invariant theories within the coset space technique. Interestingly, this requirement ensures that the virial current of conformal field theories vanishes or is a total derivative \cite{Callan:1970ze,ElShowk:2011gz}. 

Besides the above rigorous reasoning, one can arrive at the same procedure of constructing conformally invariant theories in the following way. Consider $ y^\nu $ in (\ref{Conf_Coset}) as an auxiliary field, which was introduced to ensure the applicability of the coset space technique. Then, to leave only physical fields in the theory, one should search for the Lagrangians in which $ y^\nu $ and matter fields decouple. This is just a reformulation of the requirement established in the previous paragraph. Although this reasoning is non--strict and cannot be applied directly in a spontaneously broken phase, it can be considered as an argument for the correctness of the two--orbit approach.

The extension of the area of applicability of the two--orbit approach to a spontaneously broken phase is straightforward. Namely, since conformal field theories are defined on a sphere, $ y^\nu $'s interpretations remains the same --- they are fields whose equation of motion must admit (\ref{Glue}) as a solution. This requirement follows purely from the method of induced representations and is independent from the fact whether the conformal group is spontaneously broken or not. Thus, despite that the action of SCT on the vacuum becomes non--trivial, the logic of \cite{Kharuk:2017jwe} remains valid. Another argument supporting this result is the following. As it was mentioned above, it is the condition for $ y^\nu $ to enter the Lagrangian only via total derivative that guarantees that the virial current is a total derivative. In a spontaneously broken phase the virial current must also be a total derivative, which, reversing the logic, leads to the previous condition.

Let us now discuss what this requirement implies in practice. Note that the conformal group is the maximal spacetime symmetry group that relativistic field theories are allowed to have \cite{Weinberg:2000cr,Coleman:1967ad} (of course, except for supersymmetry). This allows us to write the most general spontaneous symmetry breaking pattern including the breaking of the conformal invariance in the form
\begin{equation} \label{General_Patter}
\text{Conf}(d) \times G_{int} \rightarrow H \;,
\end{equation}
where $ G_{int} $ is some internal symmetry group and $H$ is allowed to be a mixture of spacetime and internal symmetries. Throughout the paper, we adopt the standard definition of a broken generator --- a generator is said to be broken if does not annihilate the vacuum. For the conformal group, this definition has a non--trivial consequence. To reveal it, remember that the action of SCT on quasi--primary fields (i.e., on the elements of irreducible representations) reduces to the coordinate--dependent action of dilation, Lorentz transformation and translation,
\begin{equation} \label{SCT_action}
\hat{K}_\mu \psi = \left( 2x_\mu \hat{D} - x^\nu \hat{L}_{\mu\nu} +ix_\nu x^\nu \hat{P}_\mu \right) \psi \;.
\end{equation}
This formula implies that SCT are spontaneously broken \textit{if and only if} at least one of the three generators, $ D, ~P_\mu $, and $ L_{\mu\nu} $, are broken. Indeed, if none of them are broken, then eq. (\ref{SCT_action}) implies that $ K_\mu $ are not broken as well, and vice versa. Thus, if conformal invariance is assumed to be spontaneously broken, so must be SCT (and at least one of the three generators). This observation allows one to write the coset space corresponding to (\ref{General_Patter}) in the form
\begin{equation} \label{General_Coset}
g_H = e^{iP_\mu x^\mu} e^{iK_\nu y^\nu} e^{iZ_a\xi^a} \;,
\end{equation}
where $ Z_a $ are the broken generators different from $ K_\nu $, $ \xi^a $ are the corresponding NGF, $ a $ can be a mixture of spacetime and internal indices, and it was assumed that the translations are not spontaneously broken. Provided that coset space (\ref{General_Coset}) is homogeneously reductive \cite{Ogievetsky1974}, all of the corresponding Maurer--Cartan forms except for $ \omega_H^i $,
\begin{equation}
g_H^{-1} d g_H = P_\mu \omega_P^\mu + K_\nu \omega^\nu_K + Z_a \omega^a + H_i \omega^i_H \;, 
\end{equation}
where $ H_i $ are the generators of $ H $, transform homogeneously under the action of all continuous symmetries. For a matter field $ \psi $, a homogeneously transforming 1--form reads
\begin{equation} \label{Gen_MCF}
\mathcal{D} \psi = d \psi + i \omega^i_H \hat{H}_i \psi \;,
\end{equation}
where $ \hat{H}_i $ is a representation of $ H_i $ appropriate for $ \psi $. Then, $G$-invariant Lagrangians are obtained as $H$-invariant wedge products of $ \omega_P^\mu\,, ~ \omega_K^\nu\,, ~ \omega^a_Z\,,~ \psi $ and $ \mathcal{D}\psi $ that admit (\ref{Glue}) as a solution of $ y^\nu $'s equation of motion. 

To understand which theories satisfy this criterion, note that in general case the effective Lagrangian can be split into two parts,
\begin{equation}
\mathcal{L} = \mathcal{L}_{kin} (\omega_P^\mu\,, \omega_K^\nu) + \mathcal{L}_{ph} ( \omega_P^\mu\,, \omega_K^\nu\,, \omega^a_Z\,, \psi\,, \mathcal{D}\psi ) \;,
\end{equation}
where $ \mathcal{L}_{kin} $ is a wedge product of $ \omega_P^\mu $ and $ \omega_K^\nu $ only,\footnote{One can write an analogous term, $ \mathcal{L}_{kin} $ , in the unbroken phase as well. It corresponds to $ y^\nu $'s kinetic term and always admit (\ref{Glue}) as a solution \cite{Kharuk:2017jwe}.} and $ \mathcal{L}_{ph} $ contains all other terms. As it will be explained at the end of next section, $ \mathcal{L}_{kin} $ always admits (\ref{Glue}) as a solution. Moreover, since it does not contain $ \xi^a $ and $ \psi $, all Lagrangians having the same $ \mathcal{L}_{ph} $ but different $ \mathcal{L}_{kin} $ are physically identical. Hence, without loss of generality, for the study of effective theories one can set $ \mathcal{L}_{kin} $ to zero, which will be assumed in the rest of the paper. Further, note that because of the structure of the conformal algebra, $ y^\nu $ will enter the Maurer--Cartan forms $ \omega_K^\nu $ and at least some of $ \omega_Z^a $ and 1--forms $ \mathcal{D}\psi $. This results in the appearance of the interaction terms between $ y^\nu $ and other fields. Then, since $ \xi^a $ and $ \psi $ are dynamical, the solution of $ y^\nu $'s equations of motion cannot be given by (\ref{Glue}) unless these interaction terms sum up to a total derivative. Consequently, the only allowed effective Lagrangians are those which satisfy the same criterion as in the unbroken phase.

This constitutes the sought generalization of the technique developed in \cite{Kharuk:2017jwe}. Its key finding is that the NGF for SCT does not represent fluctuations of a background solution. Instead, their presence and the way they must appear in the theories ensures the famous property of conformal field theories --- their virial current is a total derivative. 

To illustrate the application of the two--orbit approach to the construction of effective Lagrangians, consider a spontaneous breaking of the conformal group down to the Poincare one,
\begin{equation} \label{MainPattern}
\text{Conf}(d) \rightarrow SO(d) \;,
\end{equation}
From (\ref{MCF}) one reads out the covariant metric $ g_{mn} $ and the covariant derivatives of $ \pi\,,~ y^\nu $, and $ \psi $ to be
\begin{gather} \label{Inv_Metric}
g_{mn} = e^{2\pi} \delta_{mn} \,, ~~~ D_m \pi = e^{-\pi} ( \partial_m \pi + 2 y_m )  \,, \\
\label{Covar_Pi}
D_m y^\nu = e^{-2\pi} ( \partial_m y^\nu + 2y_m y^\nu - \delta^\nu_m y^2 ) \,, ~~~
\mathcal{D}_m \psi = e^{-\pi } ( \partial_m \psi + 2i y^\nu \hat{L}_{m\nu} \psi ) \;, 
\end{gather}
where $ \hat{L}^{\mu\nu} $ is a representation of $ L^{\mu\nu} $ appropriate for $ \psi $ and Latin letters denote indices which should be raised/lowered by the covariant metric. As it was discussed above, within the two-orbit approach effective Lagrangians: \textit{i}) are constructed as $SO(d)$--invaraint combinations of the covaraint derivatives, and \textit{ii}) include $ y^\nu $ only via total derivative. For example, the simplest effective Lagrangian satisfying these criteria reads
\begin{equation}
\mathcal{L} = \frac{1}{2} D_\mu \pi D^\mu \pi + D_\mu y^\mu = \frac{1}{2}e^{-2\pi}\partial_\mu \pi \partial^\mu \pi + \partial_\mu ( e^{-2\pi} y^\mu ) \;.
\end{equation}
The construction of more complicated effective Lagrangians within the two--orbit approach is problematic because of the need to ensure the second requirement. This rises the question of whether the suggested construction can be simplified. As we will show in the next section, the answer to this question is positive, and the simplified technique is nothing but the inverse Higgs constraint method. 

\section{Equivalence of the approaches}
\label{sec:3}

\subsection{The case $ d>2 $}
\label{sec:3-1}

A careful reader might have already noticed that the standard approach and the two--orbit one contradict each other. Namely, by combining equations (\ref{Glue}) and (\ref{StandardIHC}), one might see that $ \pi(x) $ must have fixed coordinate dependence, which is meaningless. This observation, in fact, demonstrates that the standard approach is not mathematically strict --- inverse Higgs constraint (\ref{StandardIHC}) is not consistent with the inversion symmetry. But then how could it happen that the standard approach turned out to be successful in all known examples? The answer is that, despite being mathematically non--strict, the standard approach allows, formally, to construct all possible effective Lagrangians. This fact is proved below by showing that any effective Lagrangian constructed in the two--orbit approach can be obtained via the standard technique as well, and vice versa. In this sense, we claim the two approaches to be equivalent. This implies that the standard technique should be considered only as a convenient tool for constructing effective Lagrangians.

As an explanatory example, the equivalence of the standard approach and the two--orbit one will be first proved for the spontaneous breaking of the conformal group down to the Poincare one, \ref{MainPattern}. After examining this case, the generalization to an arbitrary spontaneous symmetry breaking pattern will become straightforward. For the case under consideration, the coset space and Maurer--Cartan forms are given by (\ref{CosetBroken}) and (\ref{MCF}) accordingly.

First, we will show that any effective theory obtained via the two--orbit approach can be reproduced by the means of the standard technique as well. In the former approach, the effective Lagrangians are allowed to include $ y^\nu $ only via total derivative. Hence, there are no equations of motion for $ y^\nu $, and one can set $ y^\nu $ to be given by an arbitrary function of coordinates. In particular, the latter can be chosen to coincide with the expression following from the inverse Higgs constraints method, (\ref{StandardIHC}). Further, one can insert (\ref{StandardIHC}) back into the Lagrangian, which, obviously, sets $ \omega_D $ to zero. Then, since the resulting Lagrangian is conformally invariant, it must still be an $ SO(d) $--invariant combination of the left Maurer--Cartan forms,
\begin{equation}
\mathcal{L}_{ph} |_{IHC} = \mathcal{\tilde{L}} (\omega_P^\mu\,, \omega_K^\nu\,, \psi\,, \mathcal{D}\psi\,) |_{IHC} \;.
\end{equation} 
This implies that $ \mathcal{L}_{ph} $ can be rewritten as an $ SO(d) $-invariant combination of Maurer--Cartan forms (\ref{MCF}) with the imposed inverse Higgs constraint. This proves the first part of the statement.

In the reversed way, the claim will be proved by showing that a certain combination of the covariant derivatives of fields in the two--orbit technique are exactly the same as in the standard approach. The invaraint metric and covariant derivatives of fields are given by eq. (\ref{Inv_Metric}), (\ref{Covar_Pi}). Then, the crucial observation is that $ y^\nu $ enters (\ref{Covar_Pi}) linearly, which allows to use $ D_m \pi $ to eliminate $ y^\nu $ from all other covariant derivatives. For matter fields, the combination of the covariant derivatives that does not include $ y^\nu $ is
\begin{equation} \label{Covar_Matter_toIHC}
\tilde{\mathcal{D}}_m \psi = \mathcal{D}_m \psi - (i\hat{L}_{mn}\psi) D^n\pi \;.
\end{equation}
The elimination of $ y^\nu $ from $ D_m y^\nu $ is a slightly more complicated, since the latter includes $ y^\nu $'s derivative. Because $ D_m \pi $ transforms homogeneously under the action of the conformal group, one can obtain its covariant derivative in the same way as for the matter field \cite{Ogievetsky1974,Weinberg:1996kr},
\begin{equation}
\mathcal{D}_m D_n \pi = e^{-\pi} ( \partial_m D_n \pi +2iy^\lambda \hat{L}_{m\lambda} D_n \pi ) \;.
\end{equation}
Then, the modified covariant derivative of $ y^\nu $, which, in fact, does not include $ y^\nu $ and plays the role of the covariant derivative of $ \pi $, reads
\begin{equation} \label{Covar_y_toIHC}
\tilde{D}_m y_n = D_m y_n -\frac{1}{2}\mathcal{D}_m D_n \pi - \frac{1}{4}g_{mn} D_k \pi D^k \pi \;.
\end{equation} 
Since covariant derivatives (\ref{Covar_Matter_toIHC}) and (\ref{Covar_y_toIHC}) do not include $ y^\nu $, any effective Lagrangian constructed as an $ SO(d) $-invariant combination thereof automatically admits (\ref{Glue}) as a solution. Furthermore, the explicit calculation of (\ref{Covar_Matter_toIHC}) and (\ref{Covar_y_toIHC}) shows that they coincide with covariant derivatives (\ref{CovarDerMatterIHC}) and (\ref{CovarDerDilatIHC}) accordingly, used in the standard approach. Hence, any Lagrangian constructed within the standard approach can also be obtained in the two--orbit one. This finishes the proof.

Note also that the established coincidence of (\ref{Covar_Matter_toIHC}), (\ref{Covar_y_toIHC}) and (\ref{CovarDerMatterIHC}), (\ref{CovarDerDilatIHC}) respectively can be proved on the symmetry grounds. Namely, inverse Higgs constraints (\ref{StandardIHC}) establish the only relation between $ \pi $ and $ y^\nu $ that is compatible with all continuous symmetries. Hence, the elimination of $ y^\nu $ from the covariant derivatives cannot but yield the same result as if one had imposed inverse Higgs constraints. This constitutes the simplest explanation why the two approaches were found to be equivalent (in the sense discussed at the beginning of this section).

To proceed towards the general proof, first note that if at least a part of the Lorentz group is broken and $ d>2 $, then the dilations are broken as well. This follows from the fact that the scaling dimension of an operator having non--zero vacuum expectation value is bounded from below by the unitarity bound in conformal field theories,
\begin{equation} \label{UnitBound}
\Delta_\mathcal{O} > 0 \;.
\end{equation}
Consequently, any non--zero value of $ \mathcal{O} $ would also lead to the breaking of the dilation symmetry. In terms of the coset space (\ref{General_Coset}), this implies that $ D \in Z_a $, and we can write it out explicitly. The case $ d=2 $, which allows to demonstrate an important aspect of the developed approach, will be considered in the next section.

After establishing this fact, the generalization of the claim to an arbitrary spontaneous symmetry breaking pattern (\ref{General_Patter}) becomes straightforward. The reasoning above suggests that dilations are necessarily broken. Then, the proof that any effective Lagrangian constructed in the two--orbit approach can be obtained by the standard technique as well remains unchanged. To prove it in another way, note that one can always take the coset space in the form
\begin{equation} \label{Coset_With_Dilations}
g_H = e^{iP_\mu x^\mu} e^{i K_\nu y^\nu} e^{iD\pi} e^{iZ_a \xi^a} \;,
\end{equation}
where $ Z_a $ include all broken generators except for $ K_\nu $ and $ D $, and we made use of the fact that dilations are broken.  Then, because dilations commute trivially with all generators except for the translations and SCT, the dilation's Maurer--Cartan form for general spontaneous symmetry breaking pattern (\ref{General_Patter}) will be given by (\ref{MCF}). This allows to use the same reasoning as in the previous example for proving that the covariant derivatives of the fields constructed in the two--orbit approach, with eliminated $ y^\nu $, will coincide with their counterparts in the standard technique. This finishes the proof. 

The proved theorem shows that instead of following the two--orbit approach, one may use the standard inverse Higgs constraint approach. Although the letter technique is not mathematically strict (in the sense that it does not respect the inversion symmetry), it formally allows to obtain all possible effective Lagrangians. Thus, inverse Higgs constraint method should be considered as a convenient tool for constructing effective Lagrangians. 

Note that it was previously assumed that the inverse Higgs constraint should be imposed on the Maurer--Cartan form for dilations. However, if at least a part of the Lorentz group is spontaneously broken, $ y^\nu $ will also enter the Maurer--Cartan forms for the broken Lorentz transformations $ L_\alpha $. This allows one to try to eliminate $ y^\nu $ by imposing the inverse Higgs constraints
\begin{equation} \label{Alt_IHC}
\omega^\alpha_L = 0 \;,
\end{equation}
which, because of the structure of the Lorentz group, must hold for all $ \alpha $. However, as a system of equations on $ y^\nu $, (\ref{Alt_IHC}) is overdetermined. Indeed, if the dimension of the unbroken Lorentz subgroup is $ n $, then, in the coordinate form, (\ref{Alt_IHC}) is a system of
\begin{equation} \label{NumEqAltIHC}
d \times \frac{d(d-1) - n(n-1)}{2} > d
\end{equation}
equations. Because (\ref{Alt_IHC}) must hold off--shell, and because the NGF associated with the broken $ L_\alpha $ are independent, all of the appearing equations are independent as well. Hence, in $ d>2 $ they cannot be solved for $ y^\nu $, thus yielding this prescription inapplicable.

At the end of this section, let us note that the Maurer--Cartan forms for SCT for the most general spontaneous symmetry breaking pattern (\ref{General_Patter}) differ from the one obtained for the unbroken conformal group only by a factor of $ e^{-\pi} $ \cite{Kharuk:2017jwe,Volkov:1973vd}. This allows to adhere to the same kind of reasoning used in \cite{Kharuk:2017jwe} for proving that $ \mathcal{L}_{kin} $ always admit (\ref{Glue}) as a solution. Since the detailed proof of this statement is straightforward and repeats the steps made in \cite{Kharuk:2017jwe}, it will not be given in the present paper.  

\subsection{The special case $ d=2 $}
\label{sec:3-2}

Let us now discuss the special case $ d=2 $. As it will be demonstrated, in this case the use of the standard technique is problematic because of the impossibility to establish which inverse Higgs constraints should be imposed. To solve this problem, an extension of the standard technique is suggested.

We start by noticing that in two--dimensional conformal field theories\footnote{Although the conformal symmetry in $ d=2 $ is bigger than $ O(1,3) $, in this section the conformal group is understood as $ O(1,3) $.} one can imagine a vacuum solution breaking the Lorentz invariance but not the dilation symmetry.\footnote{For example, consider a vector field with quadratic kinetic term. Such field will have zero scaling dimension. Hence, its non--zero vacuum expectation value will break the Lorentz invariance but not the dilation symmetry.} In this case, (\ref{Alt_IHC}) allows to express $ y^\nu $ via $ \omega $, the NGF for the broken $ SO(2) $ symmetry. Then, within the two--orbit approach, one can use the covariant derivative of $ \omega $ to eliminate $ y^\nu $ from other covariant derivatives. Because of the symmetry restrictions, these new covariant derivatives will coincide with the ones that can be obtained in the standard approach by imposing inverse Higgs constraints (\ref{Alt_IHC}). Clearly, by repeating the reasoning from the previous section the two approaches can be proved to be equivalent in this case as well.

An interesting situation takes place when both dilation and Lorentz invariance become spontaneously broken. Now, one can eliminate $ y^\nu $ from other covariant derivatives either by using the covariant derivative of $ \omega $, or that of $ \pi $, or any combination thereof. Thus, from the perspective of the two--orbit technique, any covariant derivative obtained in this way can be used for the construction of effective Lagrangians. On the other hand, because of the symmetry restrictions, from the perspective of the standard technique this corresponds to considering inverse Higgs constraints of the form
\begin{equation}
\omega_D + \beta \omega_L = 0\;,
\end{equation}
for all possible values of $ \beta $ ($ \beta = \infty $ correspond to the condition (\ref{Alt_IHC})). Hence, in this case, the two--orbit approach is equivalent to the standard technique provided that one does not fix a particular choice of $ \beta $ but uses the covariant derivatives of fields resulting from all possible values of $ \beta $. This observation suggests the following extension of the standard technique: when it is possible to impose several inverse Higgs constraints, one should not choose between them but use the covariant derivatives resulting from all possible choices. Such prescription removes ambiguities and, as it was shown, follows from the more fundamental technique.  

Unfortunately, it is hard to provide an explicit example of a theory that would allow to illustrate the necessity of extending the standard technique in the suggested way. The reason behind this is that the scaling dimension of a field with the canonical kinetic term is zero. This observation forces one to search for exotic theories. Moreover, because the Lagrangians of conformal field theories are forbidden to include dimensionful constants, the organization of spontaneous breaking may require the consideration of quantum effects. This will require a full quantum treatement of the theory, which goes beyond the scope of the paper. Because of these facts, the discussion above will be left as a qualitative comment on the special case $ d=2 $. 

\section{Discussion and conclusion}
\label{sec:4}

The results of the previous section establish the correspondence between the two--orbit and standard techniques. The former approach is self--contained, as it follows directly from the method of induced representations. On the other hand, its application to the construction of the effective Lagrangians is complicated by the need to search for the combinations of Maurer--Cartan forms that include $ y^\nu $ only via total derivative. However, as it was shown, any effective Lagrangian obtained via the two--orbit approach can be constructed via the standard technique as well. The latter includes the imposition of inverse Higgs constraints, which does not have strict mathematical justification but allows one to construct the effective Lagrangians in a much easier way. Thus, the standard approach can be considered as a convenient tool for obtaining the effective Lagrangians, while it is the two--orbit technique that is mathematically strict and, hence, fundamental.    

The finding that the NGF for SCT always represent ``unphysical'' degrees of freedom is in agreement with the results of \cite{Watanabe:2013iia}. Namely, in \cite{Watanabe:2013iia} it was shown that if the Noether currents associated with the broken symmetries are functionally dependent, then some of the Nambu--Goldstone fields are redundant. In the conformal group, the action of the SCT reduces to the coordinate--dependent action of the translation, dilation and Lorentz transformation. This yields the Noether current for SCT to be functionally dependent on that for the latter three transformations. Moreover, because of the same property, the breaking of the SCT is always the consequence of the breaking of $ P_\mu,~ D $ or $ L_{\mu\nu} $. Hence, the NGF for SCT do not represent independent fluctuations of a background solution and always represent redundant fields.

Considering the proof of the equivalence of the techniques, it should be noted that the two approaches are found to be equivalent because of the requirement for $ y^\nu $ to enter the effective Lagrangians only via total derivative. Hence, the reasoning used in this paper does not apply for revealing the meaning of the inverse Higgs constraints in other cases \cite{Ivanov:1975zq,Low:2001bw,Ivanov:1976zq,Goon:2014ika,Ivanov:1981wn,Goon:2014paa,Nicolis:2013sga}.

Finally, the discussion of the special case $ d=2 $ showed that when it is possible to eliminate redundant degrees of freedom by imposing various inverse Higgs constraints, one should not choose between them. Instead, all of the resulting covariant derivatives can be used for the construction of the effective Lagrangians. This generalization cannot be established from the standard approach to the study of effective theories resulting from the spontaneous breaking of the conformal symmetry, but follows directly from the two--orbit technique. 

\paragraph{Acknowledgments}

The author is thankful to A. Monin and A. Shkerin for useful discussions and comments on the draft of the paper. The work was supported by the Grant 14-22-00161 of the Russian Science Foundation.

\bibliography{cftcoset}

\end{document}